\documentclass[pra,aps,twocolumn,floatfix,longbibliography,nofootinbib]{revtex4-2}
\usepackage[utf8]{inputenc}
\usepackage{natbib}
\usepackage{graphicx}
\usepackage{epsfig}
\usepackage{color}
\usepackage[tbtags]{amsmath}
\usepackage{hyperref}
\hypersetup{colorlinks,allcolors=black}
\usepackage{amssymb}
\usepackage{gensymb}
\usepackage{float}
\usepackage{amsmath}
\usepackage{tabularx,graphicx}
\usepackage{epstopdf}
\usepackage{latexsym}
\usepackage{color, colortbl}
\usepackage{xcolor}
\usepackage{psfrag}
\usepackage{bbm}
\usepackage{bm}
\usepackage{titlesec}
\usepackage{dsfont}
\usepackage{feynmp}
\usepackage{slashed}
\usepackage[utf8]{inputenc}
\usepackage{appendix}

\DeclareUnicodeCharacter{2032}{$\prime$}

\begin{document}

\title{Commensurate Antiferromagnetic Fluctuations with Small Electron Doping}

\author{Mohsen Shahbazi}

\affiliation{Department of Physics, Faculty of Science, University of Zanjan, University Blvd., 45371-38791, Zanjan, Iran}

\author{Mohammad Ali Maleki}

\affiliation{Department of Physics, Faculty of Science, University of Zanjan, University Blvd., 45371-38791, Zanjan, Iran}

\begin{abstract}

We investigate the presence of antiferromagnetic fluctuations in the longitudinal and transversal spin susceptibilities of a square lattice. The inclusion of both first and second neighbour hopping terms, along with exchange coupling, induces anti-ferromagnetic fluctuations over a finite range of  fillings in both longitudinal and transversal static spin susceptibilities. In the absence of on-site Hubbard interaction, we observe incommensurate antiferromagnetic fluctuations between the two Van Hove fillings. Beyond the second Van Hove singularity at $ n=1.03 $, commensurate antiferromagnetic fluctuations dominate in both longitudinal and transversal spin susceptibilities.  When incorporating a finite Hubbard interaction strength, $ U $, we find that the commensurate anti-ferromagnetic fluctuations are preserved in both longitudinal and transversal dressed spin susceptibilities. However anti-ferromagnetic fluctuations vanish beyond a critical value of the Hubbard coupling strength, $ U_{c} $. Despite the exchange field can be induced by a ferromagnetic substrate, we do not observe any ferromagnetic fluctuations in the system. 

\end{abstract}

\maketitle

\section{Introduction}

The study of spin fluctuations in solid-state systems has garnered significant attention due to its importance in itinerant electron magnetism \cite{moriya2012spin,thessieu1995magnetism,hasegawa1974effect} . Understanding magnetic properties such as magnetism and magnetic phase transitions is essential for both fundamental physics and technological applications \cite{takahashi2013spin,solontsov2005spin,pappas2017magnetic,paul2011magnetoelastic}. Moreover,spin fluctuations play a crucial role in understanding and explaining high-temperature superconductivity \cite{scalapino2012common,scalapino1986d,dahm2009strength} . They are also pivotal in the development of spintronics, magnetic resonance technologies, and in the refinement of theoretical models like the Heisenberg and Hubbard models \cite{gomonay2014spintronics,yazyev2008magnetic,lopez2012fluctuation,vyaselev2002superconductivity}. The interplay between electron interactions, lattice geometry, and external fields can lead to a rich variety of magnetic phenomena, making these systems a fertile ground for both theoretical and experimental research.

The $ t-t^{\prime} $ Hubbard model has been found to support a wide region of ferromagnetism around the Van Hove density \cite{hlubina1999phase}. Ferromagnetic fluctuations have also been observed in a $ t-t^{\prime} $ Rashba-Hubbard model by A. Greco et al. \cite{greco2020ferromagnetic}.  The magnetic fluctuations in the two-dimensional Hubbard model have been investigated from weak to strong coupling approximations \cite{romer2020pairing}. Furthermore, strong ferromagnetic fluctuations have been observed in a checkerboard lattice Hubbard model with electron filling \cite{pan2023strong}. The superconducting gap structures in a single-band Hubbard model in the paramagnetic limit have been studied by A. T. Rømer et al. \cite{romer2015pairing}. Additionally, a mechanism for unconventional superconductivity has been presented in the hole-doped Rashba-Hubbard model in the square lattice \cite{greco2018mechanism}.

In this work, we investigate the spin fluctuations in the $ t-t^{\prime} $ Hubbard model with an additional external exchange field. Using Wick's theorem \cite{bruus2004many}, we derive straightforward relations for both longitudinal and transversal spin susceptibilities. Our observations reveal commensurate antiferromagnetic fluctuations at low electron doping, while ferromagnetic fluctuations are absent for all possible values of filling. Our findings indicate that commensurate antiferromagnetic fluctuations persist even with a finite Hubbard interaction $ U $, highlighting the robustness of this magnetic ordering in the presence of an exchange field.

This paper is organized as follow: We begin by establishing the Hamiltonian model  and the formalism for calculating the magnetic response function (Section II). Section III delves into the calculated spin fluctuations, presenting  both bare longitudinal and transversal spin susceptibilities. Moreover, we investigate the influence of electron-electron interactions on spin susceptibilities, in Section III.  Section IV offers a concise conclusion summarizing the key findings and highlighting the originality of this work. To enhance understanding, Appendix A provides a detailed derivation of the longitudinal spin susceptibility formula. Finally, Appendix B focuses on deriving a formula for a specific element of the dressed spin susceptibility.

\section{Hamiltonian Modelling and Spin Susceptibility Characterization}
There are various frameworks available for modelling the Hamiltonian in solid-state systems. In this analysis, we employ the tight-binding framework to model a square lattice, incorporating both first- and second-neighbour hopping terms and  an external exchange field

\begin{align}\nonumber
\hat{H} = -t \sum_{\langle i,j\rangle,\sigma}  c_{i,\sigma}^{\dagger} c_{j,\sigma} + t' \sum_{\langle \langle i,j\rangle\rangle,\sigma} c_{i,\sigma}^{\dagger} c_{j,\sigma}\\
\qquad - \mu \sum_{i,\sigma}c_{i,\sigma}^{\dagger} c_{i,\sigma}+\lambda_{ex}\sum_{i,\sigma}c_{i,\sigma}^{\dagger} \sigma_{z}c_{i,\bar{\sigma}} ,
\label{eq:H0}
\end{align}
where, $c_{i}^{\dagger}$ and $c_{i}$ are the creation and annihilation operators. The Hamiltonian consists of four terms that represent different physical attributes within the system. The first term captures the nearest-neighbour hopping with a  strength of $t$ . The second term includes the next-nearest-neighbour hopping with a  strength of $t'$, which  provide a more comprehensive description of the electron dynamics. The third term stands for the chemical potential , which adjusts the electron density in the system and controls the overall energy level relative to the Fermi level.  The final term represents the exchange field, characterized by an exchange strength of $ \lambda_{ex} $, which can arise from the effect of  a ferromagnetic substrate or other magnetic influences \cite{zhao2017enhanced,shahbazi2018linear}. Notice that  $ \sigma_{z} $, stands for the z component of Pauli matrix. Moreover, $ \sigma $ and $ \bar{\sigma} $ refer to two opposite spin projections. \\

\begin{figure}
\includegraphics[trim=0 20 0 0,width=0.41\textwidth]{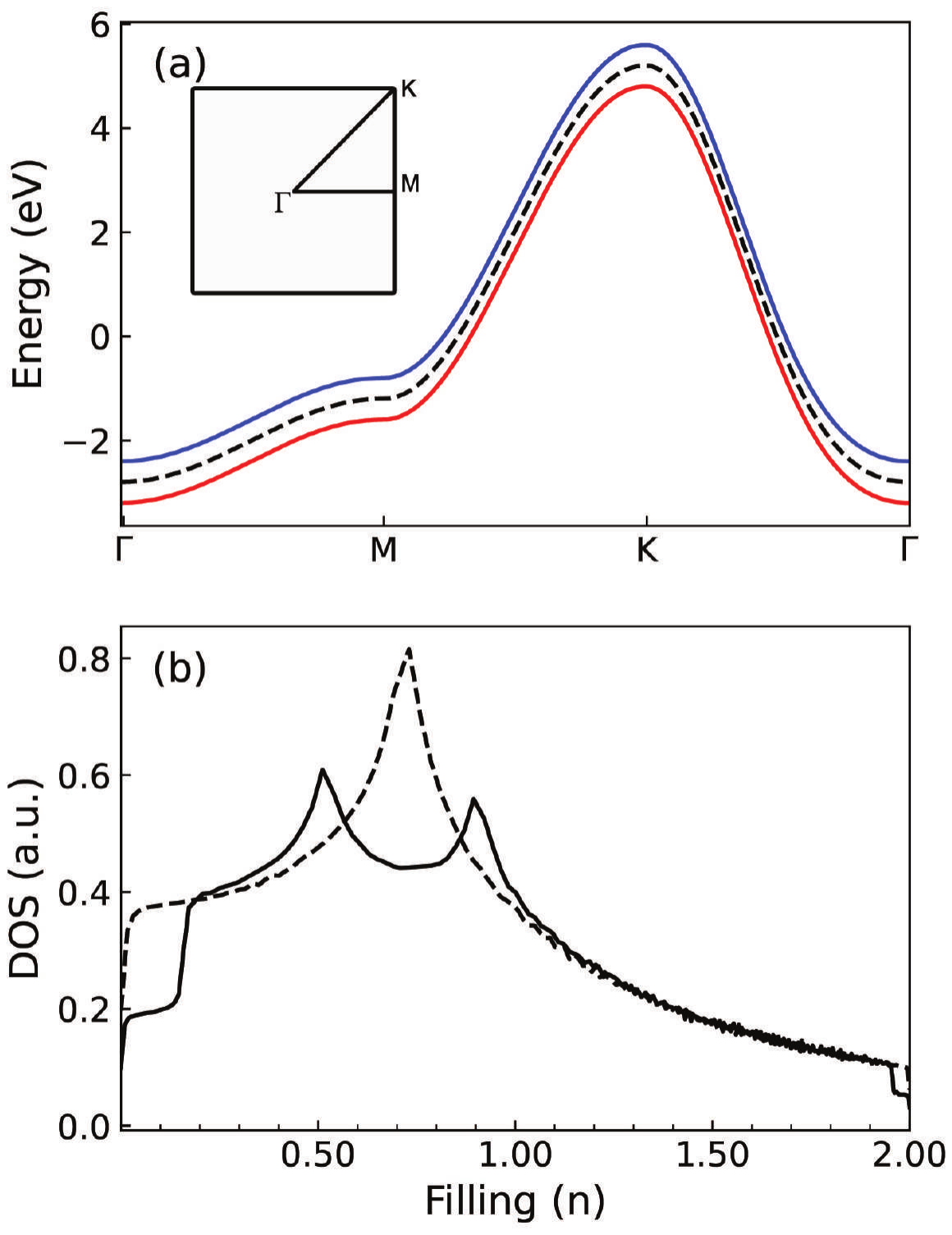}
\caption{(a) Electronic band structure of the  square lattice system, shown for two configurations. The dashed line represents the case with $t^{\prime}=0.3$ and $\lambda_{ex}=0$, while, the solid lines represent $t^{\prime}=0.3$ and $\lambda_{ex}=0.4$. Red and blue lines denote the spin-up and spin-down states, respectively. The inset illustrates the first Brillouin zone,including the high symmetry paths (b) Density of states (DOS) as a function of filling (n), for the same two configurations. The dashed line corresponds the case with  $t^{\prime}=0.3$ and $\lambda_{ex}=0$, while the solid line corresponds  $t^{\prime}=0.3$ and $\lambda_{ex}=0.4$.
}
\label{fig:band_structure_dos}
\end{figure}

The electronic band structure of the system was analyzed for two sets of parameters: first with $t' = 0.3$ and $\lambda_{\text{ex}} = 0$, and second with $t' = 0.3$ and $\lambda_{\text{ex}} = 0.4$. Note that $t = 1$ was used in all calculations. Figure \ref{fig:band_structure_dos}(a) illustrates the resulting band structures for these parameter sets. The introduction of the exchange field $ \lambda_{ex} $ causes a splitting of the energy bands for spin-up and spin-down electrons, leading to distinct density of states at the Fermi level for each spin orientation. The plot in Figure \ref{fig:band_structure_dos}(a) illustrates the metallic phase of the system, highlighting key features such as the maximum at the $ K $ point and a saddle point at the 
$ M $ point of the first Brillouin zone (FBZ).\\

Additionally, the density of states (DOS) as a function of the filling factor was computed for the same sets of initial parameters.Figure \ref{fig:band_structure_dos}(b) presents the DOS results, revealing significant details about the electronic states available at each energy level. In the absence of the exchange field ($\lambda_{ex}=0$), there is a single Van Hove singularity at a filling factor $ n_{vH}=0.74 $. This singularity corresponds to a peak in the DOS, indicating a high density of states at that energy level. When the exchange field is finite ($\lambda_{ex}=0.4$), the DOS exhibits two Van Hove singularities, located at filling factors $ n_{vH_1}=0.52 $  and  $ n_{vH_2}=0.9 $ . These additional singularities are a direct consequence of the band splitting induced by the exchange field, reflecting the separate contributions of spin-up and spin-down electrons to the overall density of states.  This behavior, as shown in Figure \ref{fig:band_structure_dos}(b), underscores the significant impact of the exchange field on the electronic properties of the system, potentially leading to interesting magnetic  phenomena. \\

So far, we have focussed the electronic properties of the system in the non-interacting limit. However, to gain a more comprehensive understanding of the material's behaviour, it is essential to consider the effects of electron-electron interactions, particularly the spin-pair interactions. It  can be effectively described using the Hubbard model. The Hamiltonian for the interacting system, incorporating the on-site Hubbard term, is given by

\begin{align}
\hat{H}_{int}=U\sum_{\textbf{k},\textbf{k}^{'},\textbf{q}}c_{\textbf{k}\uparrow}^{\dagger}c_{\textbf{k}+\textbf{q}\uparrow}c_{\textbf{k}^{'}\downarrow}^{\dagger}c_{\textbf{k}^{'}-\textbf{q}\downarrow},
\end{align}
where $ U $ represents the on-site Coulomb repulsion strength, which penalizes double occupancy of electrons at the same site.\\

To thoroughly understand the magnetic properties of the system, it is crucial to analyse the magnetic response, which is characterized by the spin susceptibility. Spin susceptibility provides a quantitative measure of how the spin configuration of the system responds to an external magnetic field. This parameter is vital for understanding various magnetic phenomena, including magnetic ordering, spin waves, and critical behaviour near phase transitions. According to the basic definition, it can be written

\begin{align}
&\chi^{zz}(\textbf{q},\tau)=\langle T_{\tau} S^{z}(\textbf{q},\tau)S^{z}(\textbf{q},0)\rangle,\\
&\chi^{+-}(\textbf{q},\tau)=\langle T_{\tau} S^{+}(\textbf{q},\tau)S^{-}(\textbf{q},0)\rangle,
\end{align}
where $ \chi^{zz} $ and $ \chi^{+-} $ refer to the longitudinal and transversal spin susceptibilities, respectively. $ \tau $  stands for imaginary time, and $ T_{\tau} $ denotes the time-ordering operator. By employing Wick’s theorem\cite{bruus2004many}, and performing a Fourier transformation with respect to time, one can express the spin susceptibilities as

\begin{align}\nonumber
\chi^{zz}(\textbf{q},i\omega_{l})=-\chi_{\uparrow\uparrow\uparrow\uparrow}(\textbf{q},i\omega_{l})-\chi_{\downarrow\downarrow\downarrow\downarrow}(\textbf{q},i\omega_{l})\\
\qquad +\chi_{\uparrow\downarrow\downarrow\uparrow}(\textbf{q},i\omega_{l})+\chi_{\downarrow\uparrow\uparrow\downarrow}(\textbf{q},i\omega_{l}),
\label{x1}
\end{align}

\begin{align}
\chi^{+-}(\textbf{q},i\omega_{l})=\chi_{\uparrow\uparrow\downarrow\downarrow}(\textbf{q},i\omega_{l})+\chi_{\downarrow\downarrow\uparrow\uparrow}(\textbf{q},i\omega_{l}),
\label{eq:t-s}
\end{align}
where $ \omega_{l}=2l\pi/\beta$ ,represents the bosonic Matsubara frequency. Note that $ \beta=\frac{1}{k_{B}T} $, where $k_{B}  $ is the Boltzmann constant and $ T $ is the temperature. More detailed calculations for longitudinal spin susceptibility are provided in Appendix A. The left-hand side terms of Equations \ref{x1} and \ref{eq:t-s} can be derived as follows

\begin{align}\nonumber
\chi_{\sigma_1\sigma_2\sigma_3\sigma_4}^{(0)}(\textbf{q},i\omega_l)=\frac{-\hbar^{2}}{4N^{2}}\sum_{\textbf{k}, ik_n}G_{\sigma_1\sigma_2}^{(0)}(\textbf{k},ik_n)\\G_{\sigma_3\sigma_4}^{(0)}(\textbf{k}+\textbf{q},ik_n+i\omega_l),
\label{eq:susceptibility}
\end{align}
where $ \sigma_i $ refers to spin-up or spin-down states. Due to the absence of spin-flip processes in our case ,  $ G_{\uparrow\downarrow}^{(0)}=G_{\downarrow\uparrow}^{(0)}=0 $. Consequently, only the following bare Green's functions are finite

\begin{align}
&G_{\uparrow\uparrow}^{(0)}(i k_n,\textbf{k})=\frac{1}{i k_{n}-E_{\textbf{k}}^{+}},
\label{eq:green_upup}
\end{align}

\begin{align}
&G_{\downarrow\downarrow}^{(0)}(i k_n,\textbf{k})=\frac{1}{i k_{n}-E_{\textbf{k}}^{-}},
\label{eq:green_dd}
\end{align}
where $  k_n=(2n+1)\pi/\beta $ refers to the fermionic Matsubara frequency. $ E_{\textbf{k}}^{+} $ and $ E_{\textbf{k}}^{-} $ are the energy dispersions of the two bands that are calculated by diagonalizing the equation (\ref{eq:H0}) as

\begin{align}\nonumber
E_{\textbf{k}}^{\pm}=&-2t\left(cos\left(k_x\right)+cos\left(k_y\right)\right)\\
&+4t^{\prime}cos\left(k_x\right)cos\left(k_y\right)-\mu \pm \lambda_{ex}.
\label{eq:energy_dispersion}
\end{align}

 There are only four finite bare spin susceptibilities, while the other possible states are zero. It should be noted that The spin susceptibility can be further decomposed into static ($ \omega_{l}=0 $) , and dynamic ($ \omega_{l}\neq 0 $)components, each revealing different aspects of the magnetic response.

\section{Results of Bare and Dressed Spin Susceptibilities}
 
To provide context, we first review certain characteristics of magnetic ordering of the system. In the square-lattice  model with a finite value for first neighbour hopping of $ t $,  and without second neighbour hopping and   exchange field ($ t^{\prime}=\lambda_{ex}=0 $), at half-filling (filling=1), the system behaves as a paramagnetic metal \cite{hirsch1985two,imada1998metal}.  The inclusion of the second-neighbour hopping term $ t^{\prime} $ modifies the shape of the Fermi surface. It can either enhance or suppress certain nesting features depending on the sign and magnitude of $ t^{\prime} $. At half-filling, the original perfect nesting vector $ Q=\left(\pi,\pi\right) $  for first-neighbour hopping is disrupted by the second-neighbor term \cite{johannes2008fermi}. This affects the tendency towards magnetic instabilities.  Without interactions, the system remains a paramagnetic metal\cite{lin1987two}. The modified band structure due to $ t^{\prime} $ does not induce spontaneous magnetic ordering by itself. Consequently, the longitudinal and transversal spin susceptibilities are equal.\\

The system with finite values of first and second neighbour hopping,$ t $ and $ t^{\prime} $, preserve time reversal symmetry and possesses spin rotational symmetry  group of $ SU(2) $.  The introduction of finite values of exchange field, $ \lambda_{ex}\neq 0 $, breaks the full $ SU(2) $ spin rotational symmetry down to $ U(1) $, as it selects a preferred direction for the spins. Moreover, the exchange field breaks time-reversal symmetry because reversing time flips the spin direction, which is not energetically favourable in the presence of the exchange field \cite{liu2022spin}. Furthermore, the system’s spin susceptibility change with the exchange field, so that the longitudinal and transversal spin susceptibilities are not equal, any more. It should be mentioned that, to calculate the static bare spin susceptibilities we perform an analytical summation over $ ik_{n} $ using the residue theorem and then we carry out a numerical summation over $ k_x $,$ k_y $ across the first Brillouin zone.\\

\begin{figure}[!ht]
\includegraphics[trim=0 20 0 0,width=0.44\textwidth]{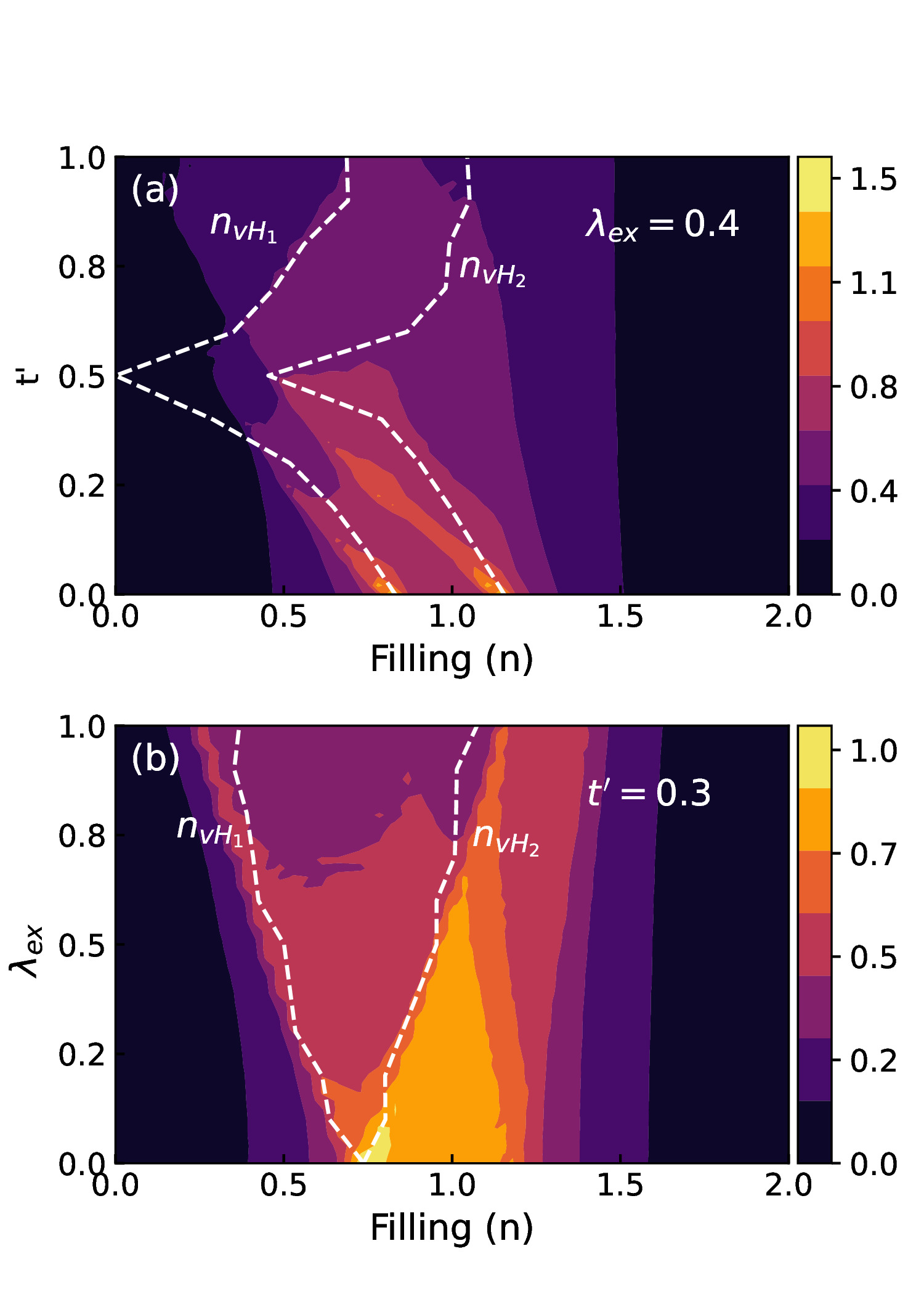}
\caption{Static bare longitudinal spin susceptibilities at $ \textbf{q}=(\pi,\pi )$ as a function of  (a)  next-nearest neighbour hopping $ t^{\prime} $ and filling, and  (b)  the exchange field $ \lambda_{ex} $ and filling. Van Hove lines are indicated by dashed lines. Note that all calculations were performed at a finite temperature of $ T=0.01 $}
\label{fig:phase_diagram}
\end{figure}

The static bare longitudinal spin susceptibility at $\textbf{q}= (\pi,\pi) $ are calculated over ranges of $ t^{\prime} $, $ \lambda_{ex} $, and $ n $ (see Figures \ref{fig:phase_diagram}(a) and \ref{fig:phase_diagram}(b)). These diagrams reveal that the longitudinal spin susceptibility exhibits large values near half-filling ($ n=1 $), particularly within the parameter ranges $0 \leq t^{\prime} \leq 0.4 $ and $0 \leq \lambda_{ex} \leq 0.4 $. As the  filling moves further away from $ n=1 $, the value of the longitudinal spin susceptibility decreases. This behaviour suggests the presence of commensurate antiferromagnetic fluctuations within these specific ranges of parameters, highlighting the potential for antiferromagnetic ordering in the system. Additionally, the Van Hove lines are calculated and plotted on these diagrams. As can be observed, these Van Hove lines cross near the peaks of longitudinal spin susceptibilities.

\begin{figure*}[!ht]
\includegraphics[trim=0 60 0 0,width=0.8\textwidth]{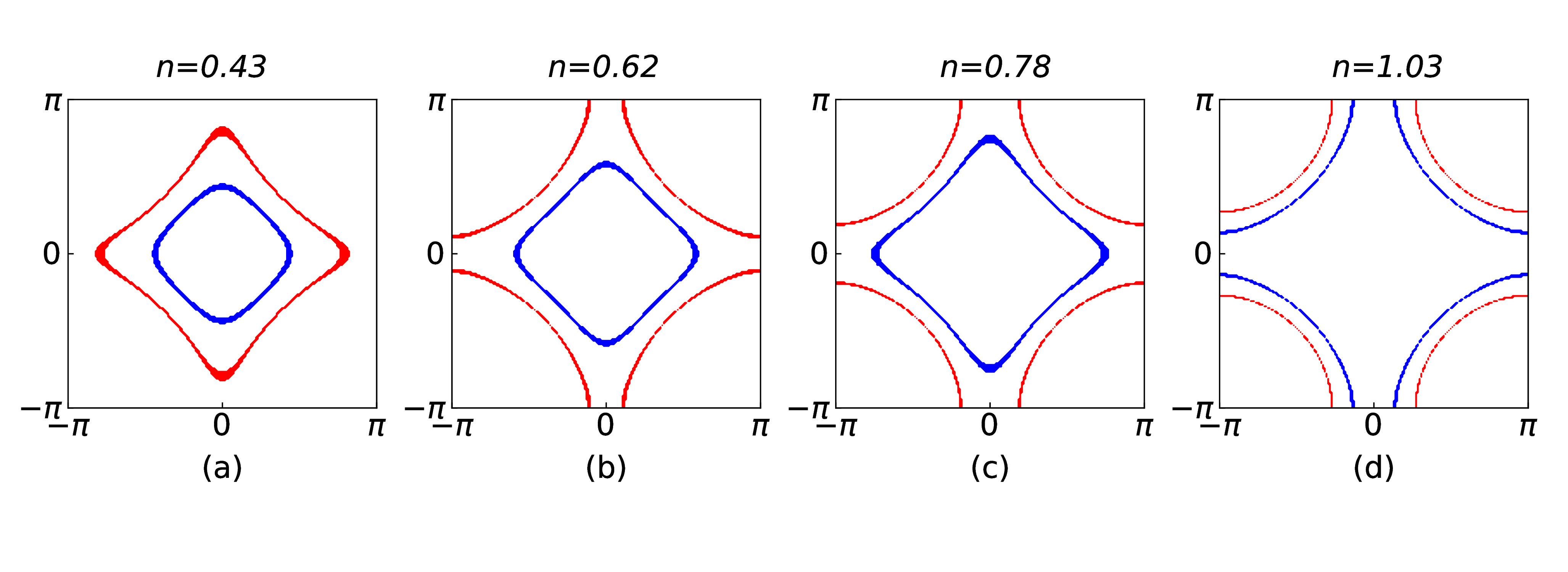}
\caption{ Fermi surfaces of the system when the parameters are set to $ t^{\prime}=0.3 $ , $ \lambda_{ex}=0.4 $, for different values of filling. Panel (a) represents the Fermi surface  with a filling less than $ n_{vH_1} $. Panels  (b) and (c) show the Fermi surfaces  with fillings between the two singularities, $ n_{vH_1} $ and $ n_{vH_2} $. Panel (d) illustrates the Fermi surface with a filling greater than $ n_{vH_2} $ }
\label{fig:fs}
\end{figure*}

\begin{figure*}[!ht]
\includegraphics[trim=0 20 0 0,width=0.8\textwidth]{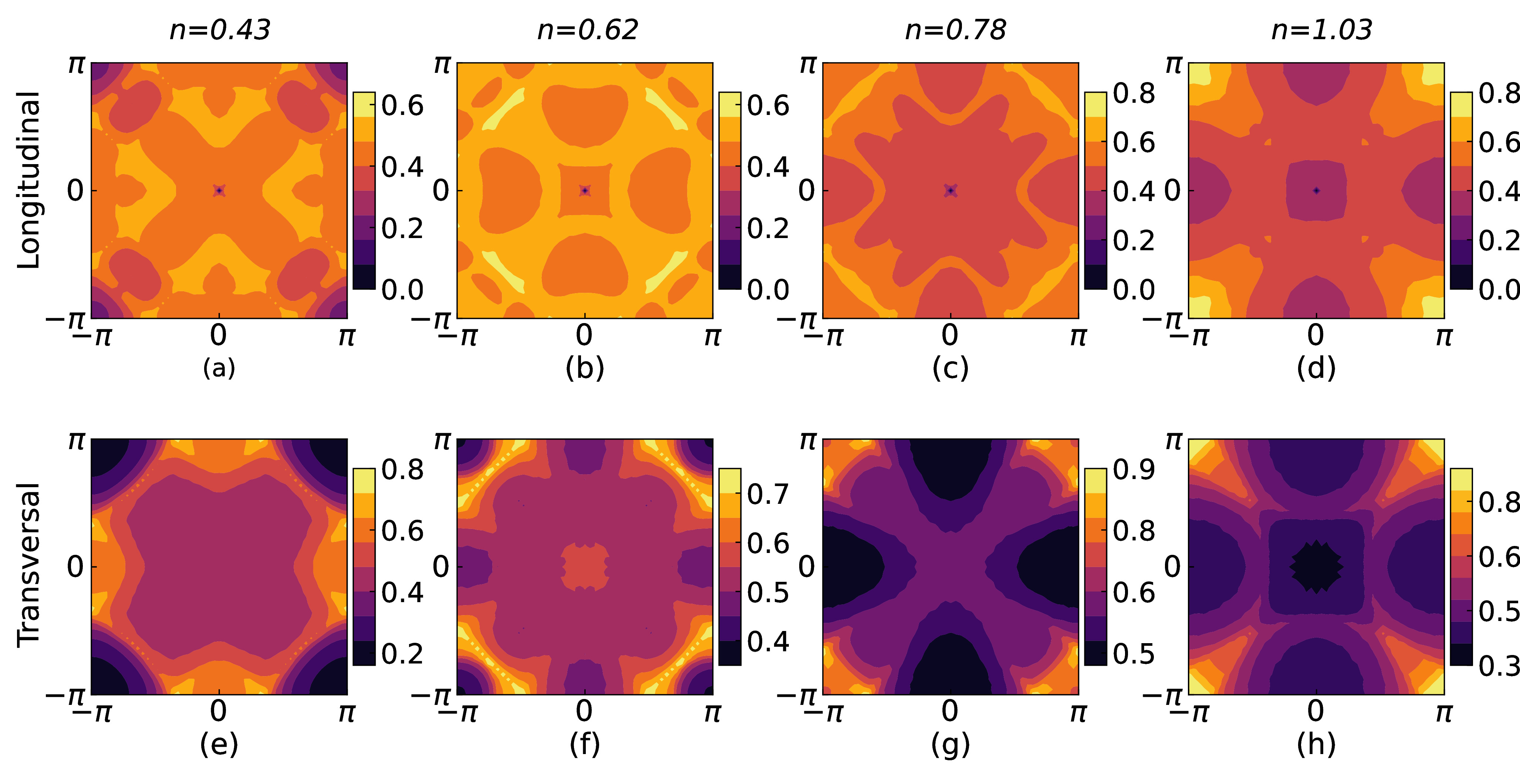}
\caption{  Static bare  spin susceptibilities with $ t^{\prime}=0.3 $, $ \lambda_{ex}=0.4 $ and  different values of filling. Panels (a-d)represent the longitudinal  and (e-h) illustrate transversal susceptibilities}
\label{fig:b_s}
\end{figure*}

It is important to explain how the magnetic ordering can be inferred from spin susceptibility plots. By analysing the peaks in the spin susceptibility, one can identify various magnetic behaviours: ferromagnetic tendencies are characterized by a peak at $ \textbf{q}=\left(0,0\right) $; commensurate antiferromagnetic behaviour is evidenced by a peak at  $ \textbf{q}=\left(\pi,\pi\right) $; and incommensurate antiferromagnetic behaviour is represented by a peak at $ \tilde{\textbf{q}} $ , which typically resides close to, but not exactly at $ \textbf{q}=\left(\pi,\pi\right) $. A factor that can significantly impact spin susceptibility is the topology of the Fermi surface \cite{romer2015pairing,hlubina1999phase}. Figures \ref{fig:fs} and \ref{fig:b_s} illustrate the Fermi surfaces and spin susceptibilities of the system under the initial conditions: $ t=1 $,$ t^{\prime}=0.3 $ and $ \lambda_{ex}=0.4 $, for different values of  filling.  For the filling with an electron-like and a hole-like Fermi surfaces, $ n=0.78 $, the longitudinal spin susceptibility demonstrates incommensurate antiferromagnetic fluctuations , see figures \ref{fig:fs}(c)and \ref{fig:b_s}(c). For a small electron doping  , $ n=1.03 $, with the two hole-like Fermi surfaces  longitudinal spin susceptibility shows commensurate antiferromagnetic fluctuations, see figures \ref{fig:fs}(D)and \ref{fig:b_s}(D). It should be mentioned that, a filling in the range $ 0 < n < 1 $ corresponds to the hole-doped regime, while a filling in the range $ 1 < n < 2 $ represents the electron-doped case.
 .Furthermore, the transversal spin susceptibilities for fillings of $ n=0.62$ and $ n=0.78$ reveal incommensurate antiferromagnetic fluctuations, as illustrated in Figures \ref{fig:b_s}(f) and \ref{fig:b_s}(g). These figures show distinct peaks at wave vectors $ \tilde{\textbf{q}} $,  which are close to but not exactly at $ \textbf{q}=\left(\pi,\pi\right) $, indicating that the system exhibits complex spin ordering rather than simple commensurate antiferromagnetic order.  This behaviour highlights the sensitivity of the transversal spin susceptibility to variations in  filling  and its role in capturing subtle changes in the magnetic structure. In contrast, the transversal spin susceptibility at $ n=1.03$ demonstrates commensurate antiferromagnetic fluctuations, as depicted in Figure \ref{fig:b_s}(h). Here, a prominent peak at $ \textbf{q}=\left(\pi,\pi\right) $ is observed, signifying a well-defined commensurate antiferromagnetic order with a modulation vector aligned with the lattice periodicity. This consistency between longitudinal and transversal susceptibilities at $ n=1.03 $  underscores the robust nature of commensurate antiferromagnetic ordering at this specific filling.

So far, we have focused on the bare spin susceptibilities $ \chi^{(0)}(\textbf{q}) $, which provide insight into the intrinsic magnetic properties of the system without considering spin interactions of pairs. However, to achieve a more accurate and comprehensive understanding of the system's magnetic behaviour, it is essential to take into account the effects of  interactions. As previously mentioned, we incorporate the  on-site Hubbard term to account for the electron-electron interactions within the system. These interaction can significantly modify the spin susceptibilities, leading to what are known as dressed spin susceptibilities $ \chi(\textbf{q})$. To investigate the impact of this interaction, we employ the random phase approximation (RPA)\cite{bruus2004many}, a powerful method that allows us to incorporate the interaction systematically. The RPA provides a framework to account for the collective effects of spin fluctuations, enhancing our understanding of the magnetic response of the system. In the following , we present the six non-zero components of the dressed spin susceptibility matrix obtained through the RPA:\\

\begin{align}
\chi_{\uparrow\uparrow\uparrow\uparrow}^{RPA}=\frac{\chi_{\uparrow\uparrow\uparrow\uparrow}^{(0)}}{1-U^{2}\chi_{\uparrow\uparrow\uparrow\uparrow}^{(0)}\chi_{\downarrow\downarrow\downarrow\downarrow}^{(0)}},
\label{eq:d_s_1111}
\end{align}

\begin{align}
\chi_{\downarrow\downarrow\downarrow\downarrow}^{RPA}=\frac{\chi_{\downarrow\downarrow\downarrow\downarrow}^{(0)}}{1-U^{2}\chi_{\downarrow\downarrow\downarrow\downarrow}^{(0)}\chi_{\uparrow\uparrow\uparrow\uparrow}^{(0)}},
\label{eq:d_s_2222}
\end{align}

\begin{align}
\chi_{\downarrow\uparrow\uparrow\downarrow}^{RPA}=\frac{U\chi_{\downarrow\downarrow\downarrow\downarrow}^{(0)}\chi_{\uparrow\uparrow\uparrow\uparrow}^{(0)}}{1-U^{2}\chi_{\downarrow\downarrow\downarrow\downarrow}^{(0)}\chi_{\uparrow\uparrow\uparrow\uparrow}^{(0)}},
\label{eq:d_s_2112}
\end{align}

\begin{align}
\chi_{\uparrow\downarrow\downarrow\uparrow}^{RPA}=\frac{U\chi_{\downarrow\downarrow\downarrow\downarrow}^{(0)}\chi_{\uparrow\uparrow\uparrow\uparrow}^{(0)}}{1-U^{2}\chi_{\downarrow\downarrow\downarrow\downarrow}^{(0)}\chi_{\uparrow\uparrow\uparrow\uparrow}^{(0)}},
\label{eq:d_s_1221}
\end{align}

\begin{align}
\chi_{\uparrow\uparrow\downarrow\downarrow}^{RPA}=\frac{\chi_{\uparrow\uparrow\downarrow\downarrow}^{(0)}}{1-U\chi_{\uparrow\uparrow\downarrow\downarrow}^{(0)}},
\label{eq:d_s_1122}
\end{align}

\begin{align}
\chi_{\downarrow\downarrow\uparrow\uparrow}^{RPA}=\frac{\chi_{\downarrow\downarrow\uparrow\uparrow}^{(0)}}{1-U\chi_{\downarrow\downarrow\uparrow\uparrow}^{(0)}}.
\label{eq:d_s_2211}
\end{align}

Further details on the calculation of the non-zero elements of s dressed spin susceptibility is provided in Appendix B.

Figure \ref{fig:d_l_s} illustrates the longitudinal dressed spin susceptibilities for various interaction strengths of $ U $, at a filling of $ n=1.03 $. As observed from Figure \ref{fig:d_l_s}, the static dressed susceptibility maintains a pronounced peak at $ \left(\pi,\pi\right) $, indicating that the commensurate antiferromagnetic fluctuation is preserved despite the inclusion of interaction effects. This persistent peak suggests that the system remains robustly in a commensurate antiferromagnetic state even as the interaction strength $ U $ is varied. The preservation of the peak at $ \left(\pi,\pi\right) $, underscores the stability of the antiferromagnetic order under the influence of electron-electron interactions. Note that this pronounced peak at $ \left(\pi,\pi\right) $  disappears at the critical point   $ U_{c}\simeq 2.5$. It highlights the dominant role of commensurate antiferromagnetic fluctuations in the magnetic response of the system at this particular filling. This finding is significant as it demonstrates that the essential magnetic characteristics are not only intrinsic to the bare system but also resilient to the perturbations introduced by the Hubbard interaction.

\begin{figure*}[!ht]
\includegraphics[trim=0 60 0 0,width=0.93\textwidth]{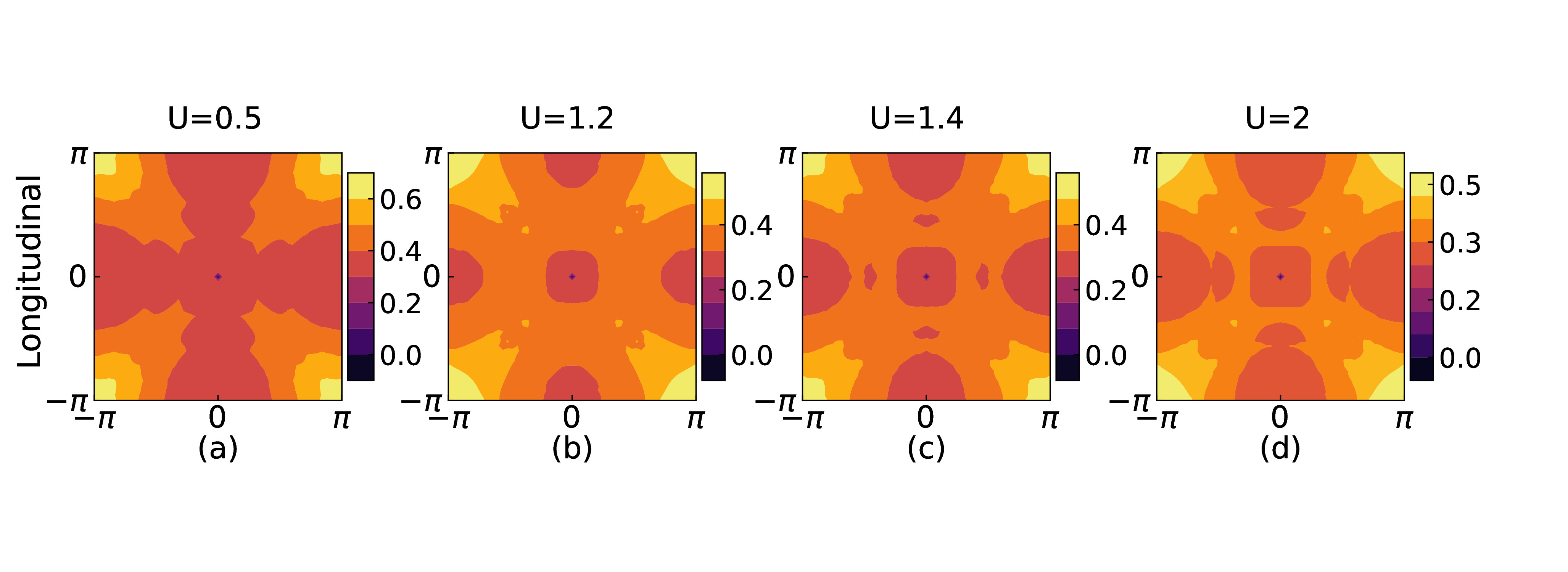}
\caption{Static longitudinal dressed spin susceptibilities at $ n=1.03 $ for different values of the on-site Hubbard coupling of U.}
\label{fig:d_l_s}
\end{figure*}

\begin{figure*}[!ht]
\includegraphics[trim=0 60 0 0,width=0.93\textwidth]{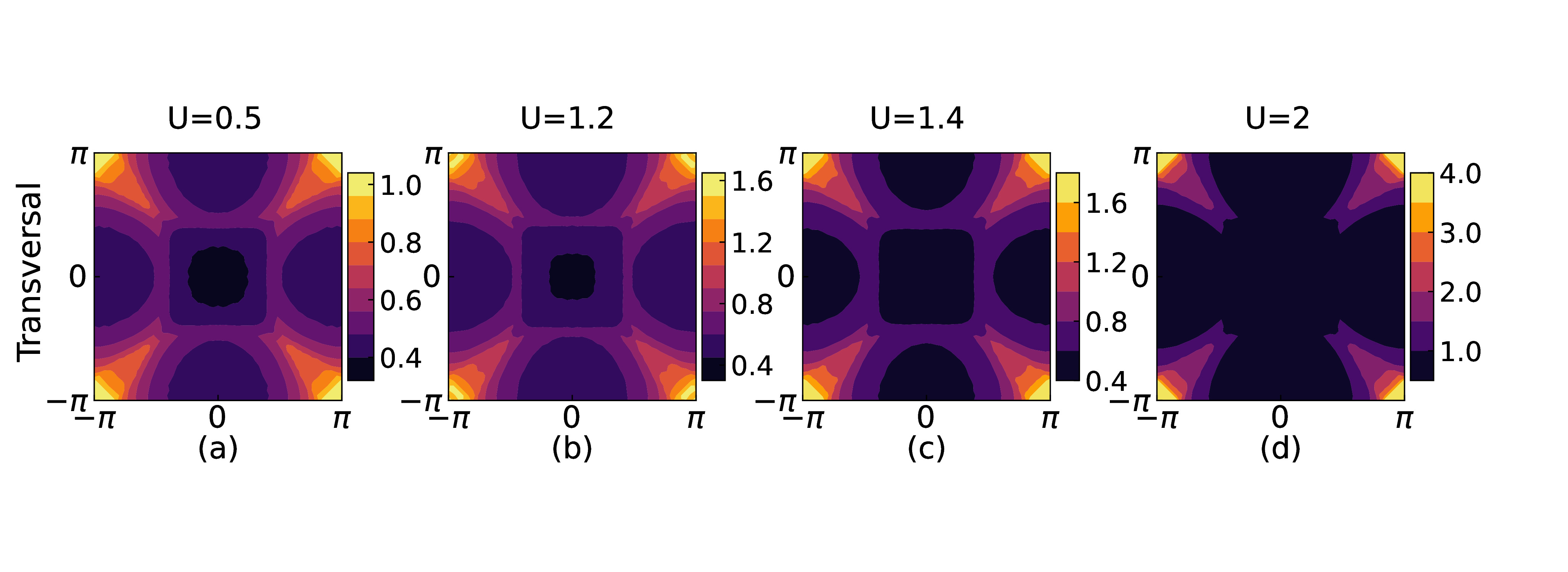}
\caption{Static transversal dressed spin susceptibilities at $ n=1.03 $ for different values of the on-site Hubbard coupling of U.}
\label{fig:d_t_s}
\end{figure*}

Figure \ref{fig:d_t_s}, presents the transversal dressed spin susceptibilities for various strengths of the Hubbard interaction $ U $ at a small electron doping of $0.03 $. The static transversal susceptibility exhibits a pronounced peak at $ \left(\pi,\pi\right) $  across different interaction strengths, indicating robust commensurate antiferromagnetic order. It is important to note that the pronounced peak at $ \left(\pi,\pi\right) $ vanishes at the critical point $ U_{c}\simeq 2.5$ .  Analyzing the dressed transversal spin susceptibilities provides a comprehensive view of the system's magnetic response. The consistency between the longitudinal and transversal susceptibilities suggests that the underlying magnetic order is intrinsically commensurate and stable in small electron doping of $0.03 $. This robustness could have significant implications for the understanding and potential manipulation of magnetic phases in correlated electron systems, offering insights into how these systems maintain their magnetic properties even in the presence of 
 interactions.

\section{Conclusion}

In summary, we have investigated the occurrence  of the  antiferromagnetic fluctuations in a system described by a Hamiltonian that includes first and second nearest neighbor hopping terms, an exchange field, and an on-site Hubbard term on a square lattice. The exchange field induces a band splitting, leading to two Van Hove singularities. In the absence of the Hubbard interaction, we find that incommensurate antiferromagnetic fluctuations dominate in the bare longitudinal and transversal spin susceptibilities, for fillings between the two Van Hove singularities. For the filling,$ n=1.03 $ , greater than the larger Van Hove filling , commensurate antiferromagnetic fluctuations become prevalent. With the inclusion of the Hubbard term, the dressed longitudinal and transversal susceptibilities reveal that commensurate antiferromagnetic fluctuations are preserved at the same filling of $ n=1.03 $. This indicates that the commensurate antiferromagnetic fluctuation  is robust at this specific electron doping. However, antiferromagnetic fluctuations vanish beyond a critical Hubbard coupling strength of $ U_{c}\simeq 2.5$. Although a ferromagnetic substrate can induce an exchange field, we do not observe any  ferromagnetic fluctuations in the system. Our findings underscore the critical role of electron interactions in shaping the magnetic properties of the system. They provide valuable insights into the stability and persistence of commensurate antiferromagnetic fluctuations at specific small electron doping, particularly in the presence of varying the strengths  on-site Coulomb repulsion .

\section*{Acknowledgement}

We express our gratitude to Dr. Andres Greco for his insightful explanations throughout the project. Additionally, Mohsen Shahbazi extends his sincere thanks to Dr. Lida Akbari for her helps.

\appendix

\section{Derivation of an expression for bare  longitudinal spin susceptibility }

In the context of condensed matter physics, the longitudinal spin susceptibility $ \chi^{zz}(q,\tau) $, is a fundamental quantity that characterizes the response of a spin system to an external perturbation aligned with the direction of the spin. Mathematically, the longitudinal  spin susceptibility is defined as

\begin{align}
\chi^{zz}(\textbf{q},\tau)=\langle T_{\tau} S^{z}(\textbf{q},\tau) S^{z}(-\textbf{q},0) \rangle,
\label{eq:S_a_1}
\end{align}  
where, $ \tau $ refers to imaginary time operator. According to the definition of the z component of the spin operator, we have

\begin{align}\nonumber
S^{z}(\textbf{q},\tau) = \frac{\hbar}{2N} \sum_{\mu,k} e^{-i\textbf{q} \cdot \mu} \left[c_{k,\uparrow}^{\dagger}(\tau)c_{k+\textbf{q},\uparrow} (\tau) - \right. \\
\left. c_{k,\downarrow}^{\dagger}(\tau)c_{k+\textbf{q},\downarrow}(\tau)\right].
\label{eq:S_a_2}
\end{align}

By substituting Equation of \ref{eq:S_a_1} into Equation \ref{eq:S_a_2} , we obtain

\begin{align}
&\chi^{zz}(\textbf{q},\tau) =  \nonumber \\
& -\frac{\hbar^{2}}{4N^{2}}  \sum_{k,k^{\prime\prime}} [ \langle T_{\tau}  c_{k,\uparrow}^{\dagger}(\tau)c_{k+\textbf{q},\uparrow}^{\dagger}(\tau)c_{k^{\prime\prime},\uparrow}(0)c_{k^{\prime\prime }-\textbf{q},\uparrow}(0)        \rangle \nonumber \\
&-\langle T_{\tau} c_{k,\uparrow}^{\dagger}(\tau)c_{k+\textbf{q},\uparrow}^{\dagger}(\tau)c_{k^{\prime\prime},\downarrow}(0)c_{k^{\prime\prime }-\textbf{q},\downarrow}(0)\rangle\nonumber \\
& -\langle T_{\tau} c_{k,\downarrow}^{\dagger}(\tau)c_{k+\textbf{q},\downarrow}^{\dagger}(\tau)c_{k^{\prime\prime},\uparrow}(0)c_{k^{\prime\prime }-\textbf{q},\uparrow}(0)\rangle \nonumber \\
& -\langle T_{\tau} c_{k,\downarrow}^{\dagger}(\tau)c_{k+\textbf{q},\downarrow}^{\dagger}(\tau)c_{k^{\prime\prime},\downarrow}(0)c_{k^{\prime\prime }-\textbf{q},\downarrow}(0)\rangle ].
\end{align}

Then, applying Wick's theorem, we can write

\begin{align}
&\chi^{zz}(\textbf{q},\tau) =  -\frac{\hbar^{2}}{4N^{2}}\sum_{k,k^{\prime\prime}}\nonumber \\
& [ -\langle T_{\tau} c_{k+\textbf{q},\uparrow}(\tau)c_{k^{\prime\prime},\uparrow}^{\dagger}(0) \rangle \langle T_{\tau} c_{k^{\prime\prime}-\textbf{q},\uparrow}(0)c_{k,\uparrow}^{\dagger}(\tau) \rangle \nonumber \\
& +\langle T_{\tau} c_{k+\textbf{q},\uparrow}(\tau)c_{k^{\prime\prime},\downarrow}^{\dagger}(0) \rangle \langle T_{\tau} c_{k^{\prime\prime}-\textbf{q},\downarrow}(0)c_{k,\uparrow}^{\dagger}(\tau) \rangle \nonumber \\
&  +\langle T_{\tau} c_{k+\textbf{q},\downarrow}(\tau)c_{k^{\prime\prime},\uparrow}^{\dagger}(0) \rangle \langle T_{\tau} c_{k^{\prime\prime}-\textbf{q},\uparrow}(0)c_{k,\downarrow}^{\dagger}(\tau) \rangle \nonumber \\
&- \langle T_{\tau} c_{k+\textbf{q},\downarrow}(\tau)c_{k^{\prime\prime},\downarrow}^{\dagger}(0) \rangle \langle T_{\tau} c_{k^{\prime\prime}-\textbf{q},\downarrow}(0)c_{k,\downarrow}^{\dagger}(\tau) \rangle].
\end{align}

So, it can be expressed as

\begin{align}
&\chi^{zz}(\textbf{q},\tau) =  -\frac{\hbar^{2}}{4N^{2}}\sum_{k,k^{\prime\prime}}\nonumber \\
& [-G_{\uparrow\uparrow}(k+\textbf{q}, \tau)G_{\uparrow\uparrow}(k, -\tau)+G_{\uparrow\downarrow}(k+\textbf{q}, \tau)G_{\downarrow\uparrow}(k, -\tau) \nonumber \\
& + G_{\downarrow\uparrow}(k+\textbf{q}, \tau)G_{\uparrow\downarrow}(k, -\tau)-G_{\downarrow\downarrow}(k+\textbf{q}, \tau)G_{\downarrow\downarrow}(k, -\tau)].
\end{align}

Therefore, by performing a Fourier transformation with respect to time, one write

\begin{align}\nonumber
\chi^{zz}(\textbf{q},i\omega_{l})=-\chi_{\uparrow\uparrow\uparrow\uparrow}(\textbf{q},i\omega_{l})-\chi_{\downarrow\downarrow\downarrow\downarrow}(\textbf{q},i\omega_{l})\\
\qquad +\chi_{\uparrow\downarrow\downarrow\uparrow}(\textbf{q},i\omega_{l})+\chi_{\downarrow\uparrow\uparrow\downarrow}(\textbf{q},i\omega_{l}),
\label{eq:l-s}
\end{align}
where $ \omega_{l}$ refers to the bosonic Matsubara frequency that is considered  zero for static susceptibility. Regarding there is no spin-flip in our case, one can express $ G_{\uparrow\downarrow}^{(0)}=G_{\downarrow\uparrow}^{(0)}=0$, then one obtain bare longitudinal susceptibility as: $\chi^{zz(0)}(\textbf{q},i\omega_{l})=-\chi_{\uparrow\uparrow\uparrow\uparrow}^{(0)}(\textbf{q},i\omega_{l})-\chi_{\downarrow\downarrow\downarrow\downarrow}^{(0)}(\textbf{q},i\omega_{l})$. Using analytical continuation of $ i\omega_{l}=\omega_l + i\eta $   , and equations \ref{eq:susceptibility}, \ref{eq:green_upup} and \ref{eq:green_dd} one write

\begin{align}\nonumber
&\chi^{zz(0)}(\textbf{q},\omega_{l})= -\frac{\hbar^{2}}{4N^{2}}\sum_{\textbf{k},ik_{n}}[\frac{1}{ik_{n}-E_{\textbf{k}}^{+}}\\\nonumber
& \frac{1}{ik_{n}-E_{\textbf{k}+\textbf{q}}^{+}+\omega_l+i\eta}+\frac{1}{ik_{n}-E_{\textbf{k}}^{-}}\\
& \frac{1}{ik_{n}-E_{\textbf{k}+\textbf{q}}^{-}+\omega_l+i\eta}],
\label{eq:chi_zz}
\end{align}
where $ ik_n $ stands for fermionic Matsubara frequency. Notice that $ E^{\pm} $ denote energy dispersions for two bands provided in the equation \ref{eq:energy_dispersion}. As can be seen, the summation is performed over both $ ik_n $ and $ \textbf{k} $. By applying the residue theorem, we perform the summation over $ ik_n $. Consequently, the summation in Equation \ref{eq:chi_zz} simplifies to a sum over the wave vector $ \textbf{k} $, as follow

\begin{align}\nonumber
&\chi^{zz(0)}(\textbf{q},\omega_{l})=\\\nonumber
&-\frac{\hbar^{2}}{4N^{2}} \sum_{\textbf{k}}[ \frac{f(E^{+}_{\textbf{k}})-f(E^{+}_{\textbf{k}+\textbf{q}}-\omega_{l}-i\eta)}{E^{+}_{\textbf{k}}-E^{+}_{\textbf{k}+\textbf{q}}+\omega_{l}+i\eta}\\
&+ \frac{f(E^{-}_{\textbf{k}})-f(E^{-}_{\textbf{k}+\textbf{q}}-\omega_{l}-i\eta)}{E^{-}_{\textbf{k}}-E^{-}_{\textbf{k}+\textbf{q}}+\omega_{l}+i\eta}],
\end{align}
where $ f(z)=(1+e^{\beta z})^{-1} $ refers to Fermi-Dirac distribution function.  It can be concluded this expression with $ \omega_l=0 $ at  $ q=(0,0) $, vanishes.

\section{ Calculation of an element of the dressed spin susceptibility matrix}

There are sixteen possible components for spin susceptibility in total, but in our system, only six of these components are non-zero due to the absence of spin-flip processes. In this section, we explain the calculation of one of the non-zero components of dressed spin susceptibility, $ \chi_{\uparrow\uparrow\uparrow\uparrow}^{RPA} $, as an example. The expansion of this component is:

\begin{align}
&\chi_{\uparrow\uparrow\uparrow\uparrow}^{RPA}=\chi_{\uparrow\uparrow\uparrow\uparrow}^{(0)}+
\chi_{\uparrow\uparrow\uparrow\uparrow}^{(0)}U\chi_{\uparrow\uparrow\uparrow\uparrow}^{(0)}U\chi_{\uparrow\uparrow\uparrow\uparrow}^{(0)} \nonumber \\
&+\chi_{\uparrow\uparrow\uparrow\uparrow}^{(0)}U\chi_{\uparrow\uparrow\uparrow\uparrow}^{(0)}U\chi_{\uparrow\uparrow\uparrow\uparrow}^{(0)}U\chi_{\uparrow\uparrow\uparrow\uparrow}^{(0)}U\chi_{\uparrow\uparrow\uparrow\uparrow}^{(0)} \nonumber \\
&+...
\end{align}

\begin{figure}[H]
\includegraphics[width=0.44\textwidth]{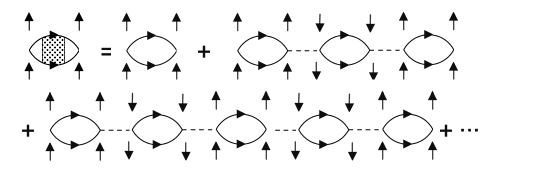}
\caption{A detailed schematic of the diagrammatic expansion for $ \chi_{\uparrow\uparrow\uparrow\uparrow}^{RPA} $. Solid lines with arrows illustrate propagators. The spin orientations are shown by  arrows out of bubbles.Dashed lines indicate the interaction term $ U $ }
\label{fig:Feynman_1}
\end{figure}

In the context of Feynman diagrams, spin susceptibilities are represented by specific diagrammatic elements: bubbles and ladders. These diagrams help visualize and calculate the modifications to the susceptibilities due to interactions. The corresponding  RPA expansion of  $ \chi_{\uparrow\uparrow\uparrow\uparrow}^{RPA} $  is shown in the Figure \ref{fig:Feynman_1}

\begin{figure}[H]
\includegraphics[width=0.44\textwidth]{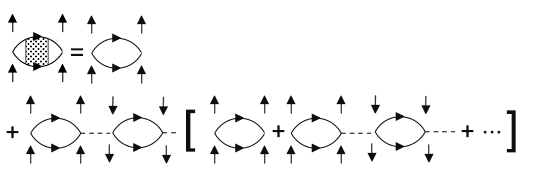}
\caption{Factoring of the expansion of the  $ \chi_{\uparrow\uparrow\uparrow\uparrow}^{RPA} $  }
\label{fig:Feynman_2}
\end{figure}

By  factoring, we arrive at the diagram shown in Figure \ref{fig:Feynman_2}

\begin{figure}[H]
\includegraphics[width=0.36\textwidth]{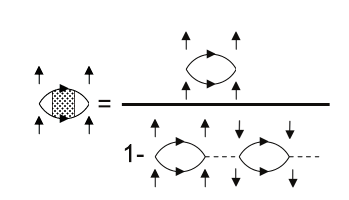}
\caption{Schematic representation of the Feynman diagram for the derived $ \chi_{\uparrow\uparrow\uparrow\uparrow}^{RPA} $}
\label{fig:Feynma_3}
\end{figure}

After renormalization, the result can be seen in Figure \ref{fig:Feynma_3}

Therefore, one can write as the corresponding formula as

\begin{align}
\chi_{\uparrow\uparrow\uparrow\uparrow}^{RPA}=\frac{\chi_{\uparrow\uparrow\uparrow\uparrow}^{(0)}}{1-U^{2}\chi_{\uparrow\uparrow\uparrow\uparrow}^{(0)}\chi_{\downarrow\downarrow\downarrow\downarrow}^{(0)}}.
\label{eq:d_s_1111}
\end{align}

The rest of the  non-zero components of dressed spin susceptibilities can be obtained using a similar procedure.

\end{document}